\documentclass{aa}  

\usepackage{natbib,graphicx,amssymb,amsmath}
\bibpunct{(}{)}{;}{a}{}{,} 
\newcommand{\kms}{$\rm km\,s^{-1}$} 
\newcommand{\mc}{\multicolumn} 
\newcommand{\tfm}{\tablefootmark}
\newcommand{\tft}{\tablefoottext}

\DeclareGraphicsExtensions{.pdf}
\usepackage{color}

\begin{document}

\title{The unbearable opaqueness of Arp220}
\titlerunning{The unbearable opaqueness of Arp220}
\author{
  S. Mart\'in \inst{\ref{inst1},\ref{inst2},\ref{inst3}}
  \and
  S. Aalto \inst{\ref{inst4}}
  \and
  K. Sakamoto \inst{\ref{inst5}}
  \and
  E. Gonz\'alez-Alfonso \inst{\ref{inst6}}
  \and
  S. Muller \inst{\ref{inst4}}
  \and
  C. Henkel\inst{\ref{inst7},\ref{inst8}}
  \and
  S. Garc\'ia-Burillo\inst{\ref{inst9}}
  \and
  R. Aladro \inst{\ref{inst1},\ref{inst4}}
  \and
  F. Costagliola \inst{\ref{inst4},\ref{inst10}}
  \and
  N. Harada\inst{\ref{inst5}}
  \and
  M. Krips\inst{\ref{inst3}}
  \and
  J. Mart\'in-Pintado\inst{\ref{inst11}}
  \and
  S. M\"uhle\inst{\ref{inst12}}
  \and
  P. van der Werf\inst{\ref{inst13}}
  \and
  S. Viti\inst{\ref{inst14}}
}
\institute{
  European Southern Observatory, Alonso de C\'ordova 3107, Vitacura, Santiago, Chile \email{smartin@eso.org}\label{inst1}
  \and 
  Joint ALMA Observatory, Alonso de C\'ordova 3107, Vitacura, Santiago, Chile\label{inst2}
  \and
  Institut de Radio Astronomie Millim\'etrique, 300 rue de la Piscine, Dom. Univ., 38406, St. Martin d'H\`eres, France \label{inst3}
  \and
  Department of Earth and Space Sciences, Chalmers University of Technology, Onsala Space Observatory, 439 92 Onsala, Sweden\label{inst4}
  \and
  Institute of Astronomy and Astrophysics, Academia Sinica, PO Box 23-141, 10617 Taipei, Taiwan\label{inst5}
  \and
  Universidad de Alcal\'a de Henares, Departamento de F\'isica y Matem\'aticas, Campus Universitario, 28871 Alcal\'a de Henares, Madrid, Spain\label{inst6}
  \and
  Max-Planck-Institut f{\"u}r Radioastronomie, Auf dem H{\"u}gel 69, 53121 Bonn, Germany\label{inst7}
  \and
  Astronomy Department, King Abdulaziz University, P.O. Box 80203, Jeddah 21589, Saudi Arabia\label{inst8}
  \and
  Observatorio Astron\'omico Nacional (OAN)-Observatorio de Madrid, Alfonso XII 3, 28014 Madrid, Spain\label{inst9}
  \and
  Osservatorio di Radioastronomia (ORA-INAF), Italian ALMA Regional Centre, c/o CNR, Via Gobetti 101, 40129 Bologna, Italia \label{inst10}
  \and
  Centro de Astrobiolog\'ia (CSIC-INTA), Ctra. de Torrej\'on a Ajalvir km 4, 28850 Torrej\'on de Ardoz, Madrid, Spain \label{inst11}
  \and
  Argelander-Institut f\"ur Astronomie, Auf dem H\"ugel 71, 53121 Bonn, Germany \label{inst12}
  \and
  Leiden Observatory, Leiden University, 2300 RA, Leiden, The Netherlands \label{inst13}
  \and
  Department of Physics and Astronomy, UCL, Gower St., London, WC1E 6BT, UK \label{inst14}
}
  

\abstract
{
The origin of the enormous luminosities of the two opaque nuclei of Arp~220, the prototypical ultra-luminous infrared galaxy, remains
a mystery
because we lack observational tools to 
explore the innermost regions around the nuclei.
}
{
We explore the potential of imaging vibrationally excited molecular emission at high angular resolution to better understand the morphology and physical structure of the dense gas in Arp~220 and to gain insight into the nature of the nuclear powering sources.
}
{
The Atacama Large Millimeter/submillimeter Array (ALMA) provided simultaneous observations of HCN, HCO$^+$, and vibrationally excited HCN $v_2=1f$ emission. Their $J=4-3$ and $3-2$ transitions were observed at a matching resolution of $\sim0.5''$, which allows us to isolate the emission from the two nuclei.
}
{
The HCN and HCO$^+$ lines within the ground-vibrational state poorly describe the central $\sim100$~pc region around the nuclei because there are strong
effects of cool absorbing gas in the foreground and severe line blending that is due to the prolific molecular emission of Arp~220.
Vibrationally excited emission of HCN is detected in both nuclei with a very high ratio relative to the total $L_{FIR}$, higher than in any other observed galaxy and well above what is observed in Galactic hot cores. HCN $v_2=1f$ is observed to be marginally resolved in $\sim60\times50$~pc regions inside the dusty $\sim100$~pc sized nuclear cores. Its emission is centered on our derived individual nuclear velocities based on HCO$^+$ emission ($V_{WN}=5342\pm4$ and $V_{EN}=5454\pm8$~\kms, for the western and eastern nucleus, respectively). 
With virial masses within $r\sim25-30$~pc based on the HCN~$v_2=1f$ line widths, we estimate gas surface densities (gas fraction $f_g=0.1$) of $3\pm0.3\times10^4~M_\odot~\rm pc^{-2}$ (WN) and $1.1\pm0.1\times10^4~M_\odot~\rm pc^{-2}$ (EN).
The $4-3/3-2$ flux density ratio could be consistent with optically thick emission, which would further constrain the size of the emitting region to $>15$~pc (EN) and $>22$~pc (WN).
The absorption systems that may hide up to $70\%$ of the HCN and HCO$^+$ emission are found at velocities of $-50$~\kms~(EN) and $6$, $-140$, and $-575$~\kms (WN) relative to velocities of the nuclei.
Blueshifted absorptions are the evidence of outflowing motions from both nuclei.
}
{
Although vibrationally excited molecular transitions could also be affected by opacity, they may be our best tool to peer into the central few tens of parsecs around compact obscured nuclei like those of Arp~220.
The bright vibrational emission implies the existence of a hot dust region radiatively pumping these transitions. We find evidence of a strong temperature gradient that would be responsible for both the HCN $v_2$ pumping and the absorbed profiles from the vibrational ground state as a result of both continuum and self-absorption by cooler foreground gas.
}

\maketitle

\section{Introduction}
\label{sect.intro}
The study of the innermost regions around the nuclei of luminous and ultra-luminous infrared galaxies (U-LIRGs) can be hindered by
extreme obscuration ($A_v>1000$) that is due to the large amounts of intervening dust.
This obscuration can hide the central power source at most wavelengths.
Molecular emission has been exploited as an alternative way to circumvent this limitation as its rotational transitions spanning the mm and
submm wavelength ranges are less affected by obscuration. However, even using these emission lines we  may have difficulties to peer into the most obscured and
compact nuclei such as those found in Arp~220, the closest prototypical ULIRG. Therefore, we
need to explore new tools and/or tracers to study the central regions of such elusive objects. 

The two nuclei within the advanced merger system of Arp~220, separated by $\sim1"$ ($\sim370$~pc for $D=76$~Mpc), are surrounded by two high column density counterrotating gas disks that heavily obscure their inner nuclear regions \citep{Sakamoto1999}. These nuclei are optically thick outside of the $5-350$~GHz frequency range as a result of free-free and dust absorption at low and high frequencies \citep{Barcos-Munoz2015}.
Although they have been the target of studies covering all available wavelengths for decades, the origin of the $\sim 10^{12}L_\odot$ stemming from Arp~220 remains an unanswered question, with arguments suggesting an active galactic nucleus (AGN) and/or a starburst origin \citep[see references in][]{Wilson2014,Tunnard2015,Barcos-Munoz2015}.
Even the many molecular transitions that have been detected and imaged at high resolution \citep[][and references therein]{Mart'in2011} have not
been able to provide a definitive answer.
Nevertheless, at the highest resolution achieved with mm and submm interferometry, 
Arp~220 reveals new unexpected features such as the P-Cygni profiles revealed by HCO$^+$ and SiO \citep{Sakamoto2009,Tunnard2015},
and possibly maser HNC emission \citep{Aalto2009}.

The recent advent of ALMA therefore led different groups to redirect their focus back to Arp~220 in an ultimate effort
to reveal its nuclear power source \citep{Wilson2014,Rangwala2015,Scoville2015}.
However, a new obstacle was found in the form of line absorption features that erase the molecular emission from the very central regions.
This absorption does not only affect the CO emission \citep{Rangwala2015}, but also dense gas tracers such as HCN, HCO$^+$, and CS
\citep{Scoville2015,Aalto2015a}, as well as their rarer isotopic substitutions \citep{Tunnard2015}, which were so far always regarded as a solution for optical thickness affecting the main isotopologs.

Thus it becomes essential to find and target species and/or transitions that are only created and excited within the nuclear region and are absent in the outer and colder gas structures, which veil our capacity to study these elusive regions.

During the past few years, a number of studies have revealed vibrationally excited transitions of HCN and HC$_3$N in extragalactic sources. This emission has only been detected toward some nearby obscured nuclei, namely the LIRG NGC~4418, and the ULIRGs Arp~220, IRAS20155-4250, and Mrk~231
\citep{Sakamoto2010,Costagliola2010,Costagliola2013,Costagliola2015, Mart'in2011,Imanishi2013,Aalto2015}.
Vibrationally excited levels can be efficiently pumped through mid-IR photons from hot gas with dust temperatures $T_d>100$~K \citep{Sakamoto2010,Aalto2015a}.
These levels have therefore been claimed to be a powerful tool to isolate the innermost regions associated with these heavily obscured nuclei.

In this paper we present the first spatially resolved imaging of the vibrationally excited emission in the two nuclei of Arp~220. These observations also clearly locate the absorption complexes affecting the dense gas tracers HCN and HCO$^+$ in their vibrational ground-state transitions.
Together with the detected continuum, we study the physical structure of the two nuclei.

\section{Observations}
The observations were carried out with ALMA in its Cycle 1 as part of a spectral line survey of Arp~220 (Mart\'in et al. in preparation).
In this paper we report the measurements covering HCN and HCO$^+$ in their $J=3-2$ (261.152~GHz and 262.794~GHz) and $J=4-3$ (348.194~GHz and 350.383~GHz) transitions, as well as the corresponding $v_2=1$ transitions of HCN (see Sect.~\ref{sect.vibresults} for details).
These sky frequencies are redshifted, adopting $z=0.018126$.

Band 6 observations were carried out on July 8,$^{}$ 2014, with 31 antennas. The correlator was configured on each sideband with $2\times1.875$~GHz spectral windows overlapping by 100~MHz.
The resulting two 3.65~GHz bands were centered on the sky frequencies 247.1 and 262.5~GHz.
J1550+0527, Titan, and J1516+1932 were used as bandpass, flux, and gain calibrators.
The median system temperature was 80~K during the observations. The spectral resolution of  0.488~MHz was degraded by 48 channels to a velocity resolution of $\sim 27-28$~\kms.
The short snapshot with 2.5~minutes of on-source integration resulted in an rms of 1.1-1.3~mJy~beam$^{-1}$ at the smoothed resolution.

Band 7 observations were carried out in two separate snapshots on April 27$^{}$ (34 antennas) and May 27,$^{}$ 2014 (31 antennas).
The correlator was configured similarly to band 6, but the two bands were centered on the sky frequencies 336.8 and 348.7~GHz.
Calibrators were the same as those for band 6. 
Median system temperatures were 210 and 140~K at 336.8 and 348.7
GHz, respectively. The resolution was degraded by 60 channels to a velocity resolution of $\sim 25-26$~\kms.
The total on-source integration of 7.2~minutes resulted in an rms of $\sim1$~mJy~beam$^{-1}$ per smoothed channel.

In both bands, flux calibration used the Butler-JPL-Horizons 2012 models. Regions affected by emission line in the Titan data were excluded. Flux accuracy is expected to be better than $10-15\%$.

The phase center of the observations was $\alpha_{J2000}=15^{\rm h}34^{\rm m}57\fs271$, $\delta_{J2000}=23^\circ30'10\farcs480$, which is offset by $\sim1''$ in declination from the position of the Arp~220
nuclei (Sect.~\ref{sect.continuum}).
The similar {\em uv}-coverage in both bands resulted in matching $\sim 0.5''$ synthesized beams (Table~\ref{tab.continuum}).


Continuum subtraction was performed in the {\em uv} visibilities.
Line contribution to the continuum emission has been shown to be a problem in Arp~220 \citep{Mart'in2011}, where the spectra are confusion limited because of both the prolific molecular emission and broad line widths. This becomes even more of a problem at high frequency, as has also been observed
toward the LIRG NGC~4418 \citep{Costagliola2015},
and it hinders finding line-free spectral regions to extract continuum information.
Although the spectral lines presented in this paper were observed in the upper sideband of the two setups, we used the continuum windows available in both bands to better account for the spectral index ($\alpha\sim3$) at these frequencies.
For each of the four $3.65$~GHz bands, we selected regions that we estimated to be free of molecular emission based on both visual inspection and spectroscopic catalogs. The aggregated (non-continuous) bandwidth used to estimate the continuum emission
is indicated in Table~\ref{tab.continuum}. The frequencies indicated for each continuum measurement represent the average frequency of the ranges we used, and therefore do not match the center of the observed bands.
About $15-25\%$ of the total bandwidth was usable for continuum subtraction (except for the lower sideband of band 6 where only $5\%$ of the band was estimated to be line free). 
A first-order linear fit to the continuum emission was used individually for bands 6 and 7.

The data were calibrated and deconvolved using CASA release 4.2.1 \citep{McMullin2007}.

\section{Results}
\label{sec.results}
\subsection{Continuum emission}
\label{sect.continuum}

\begin{figure*}
\centering
\includegraphics[width=\linewidth]{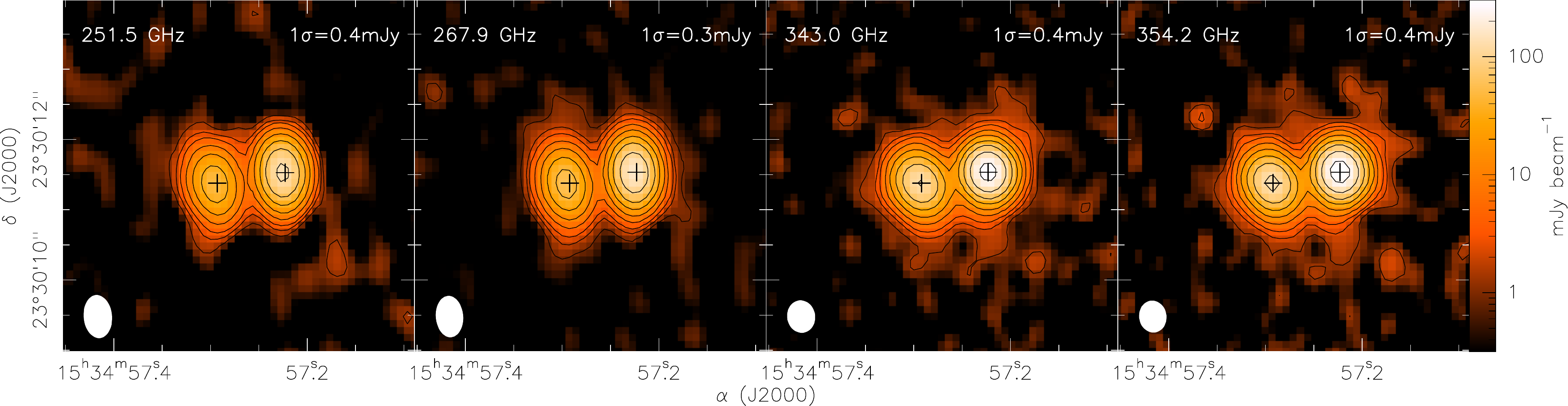}
\caption{Continuum images of the four observed bands. Central rest frequencies are shown in the upper left corner. Contours represent the 3, 6, 10, 20, 40, 80, 150, 300, and 600~$\sigma$ levels, where the $\sigma$ for each image is indicated
in the upper right corner. As a reference, the fitted positions for the two nuclei at 354.2 GHz are indicated as crosses.
 \label{fig.continuum}}
\end{figure*}

In Table~\ref{tab.continuum} we present the total integrated continuum fluxes derived for each individual observed band. Continuum images of each band are shown in Fig.~\ref{fig.continuum}.

\begin{table}
\caption{Integrated measured continuum flux densities \label{tab.continuum} }
\scriptsize
\begin{tabular}{l c c c c c}
\hline
Sky Freq.    &    Synt. Beam\tfm{a}       & Band\tfm{b}  &  rms     &     Flux\tfm{c}         &    ext.\tfm{d}      \\
(GHz)        &  ($''\times''~(^\circ)$)  & (MHz) &  (mJy)   &     (mJy)         &    frac.         \\ 
\hline
\hline
247.0        &   $0.62\times0.40~(5.5)$   & 187   &  0.4    &      225 (15)     &     $3\%$        \\
263.1        &   $0.59\times0.39~(4.2)$   & 890   &  0.3    &      286 (12)     &     $6\%$        \\
336.9        &   $0.46\times0.40~(8.7)$   & 790   &  0.4    &      570 (20)     &     $8\%$        \\
347.9        &   $0.46\times0.39~(8.6)$   & 615   &  0.4    &      630 (20)     &     $8\%$        \\
\hline
\end{tabular}
\tablefoot{
\tft{a}{Synthesized beam size in arcseconds and P.A. in degrees.}
\tft{b}{Aggregated bandwidth used to estimate and subtract the continuum emission in the {\em uv}-plane.}
\tft{c}{Flux density integrated over the central $4''$.}
\tft{d}{Fraction of the total continuum flux density found to be extended outside the nuclear concentrations.}
}
\end{table}

The good signal-to-noise ratio of the observations allowed fitting a double Gaussian in the image plane to the two nuclei.
Results are presented in Table~\ref{tab.continuum2}.
The relative positions of the nuclei can be determined with an accuracy of $\sim 0.02''$.
The positions derived for all bands agree within the uncertainties. 
Based on the measurement in our band 7 data, we derive positions of the continuum peaks at
$\alpha_{J2000}=15^{\rm h}34^{\rm m}57.223^s$, $\delta_{J2000}=23^\circ30'11.53''$ for the western nucleus (WN) and 
$\alpha_{J2000}=15^{\rm h}34^{\rm m}57.293^s$, $\delta_{J2000}=23^\circ30'11.38''$ for the eastern nucleus (EN).
These positions agree with the many estimates in the literature \citep[see Fig. 3 in][]{Scoville2015}.
The measured source sizes in band 6 of the compact components agree well with the sizes derived by \citet{Wilson2014}, but they are still marginally resolved at our resolution.

Continuum emission is significantly detected outside the compact structures around the nuclei. As an estimate of the amount of extended emission compared to that enclosed in the nuclei, Table~\ref{tab.continuum} shows the ratio between the overall emission integrated within the central $\sim4''$ ($\sim1.5$~kpc) and
that extracted from the Gaussian fits. We derive that $\sim8\%$ of the emission is extended in band 7.
Our continuum measurements agree with those in the literature \citep[see references in][]{Matsushita2009}. As discussed by \citet{Sakamoto2008}, single-dish observations result in a total flux density at 345~GHz ranging from $0.7-0.8$~Jy. While \citet{Sakamoto2008} considered that the CO contribution could add up to $11\%$ of the bolometric observations, we now know that overall molecular emission could account for up to $>20\%$ in Arp~220 \citep{Mart'in2011}. Thus we can conclude that the our ALMA observations recover all the continuum flux density.
At 247~GHz, extended emission only accounts for $3\%$ of the total emission, which is due to the lower sensitivity in this band as a result of the limited line-free region available for continuum measurement (Table~\ref{tab.continuum}).

\begin{table*}
\caption{Gaussian fit to the continuum emission in the two nuclei \label{tab.continuum2}}
\begin{tabular}{c c c c c c c c }
\hline
Sky Freq.   &   Rest Freq &  $\alpha_{J2000},\delta_{J2000}$        &  Convolved Size           &   Deconv. Size             &   Peak Flux       & Peak T$\rm_b$~\tfm{a} & Int. Flux \\
(GHz)       &   (GHz)     &  $15^{\rm h}34^{\rm m}, 23^\circ30'$    &($''\times''~(^\circ))$)   & ($''\times''~(^\circ))$)   & (mJy~beam$^{-1}$) &   (K)                    &   (mJy)   \\
\hline
\hline
            &        \mc{5}{c}{\large \bf Western Nucleus}  \\
247.0       &   251.5     &      57.226$^s$, 11.52$''$              &   $0.65\times0.44~(5)$  &   $0.22\times0.18~(1)$     & 132.0 (1.0)  & 80            & 153.2 (1.2)  \\  
263.1       &   267.9     &      57.226$^s$, 11.52$''$              &   $0.63\times0.43~(4)$  &   $0.22\times0.19~(2)$     & 159.7 (1.2)  & 80            & 189.3 (1.5)  \\  
336.9       &   343.0     &      57.223$^s$, 11.53$''$              &   $0.54\times0.46~(155)$&   $0.32\times0.16~(133)$   & 259.0 (2.9)  & 70            & 348.4 (3.8)  \\  
347.9       &   354.2     &      57.223$^s$, 11.53$''$              &   $0.53\times0.46~(155)$&   $0.32\times0.16~(132)$   & 283.1 (3.1)  & 80            & 384.2 (4.2)  \\  
             &        \mc{5}{c}{\large \bf Eastern Nucleus}  \\
247.0       &   251.5     &       57.296$^s$, 11.36$''$              &   $0.70\times0.47(12)$ &   $0.36\times0.22(32)$      &  48 (1.0)    & 17             &  64.0 (1.3)  \\ 
263.1       &   267.9     &       57.296$^s$, 11.37$''$              &   $0.69\times0.46(12)$ &   $0.38\times0.23(31)$      &  58 (1.2)    & 16             &  80.5 (1.7)  \\ 
336.9       &   343.0     &       57.293$^s$, 11.38$''$              &   $0.59\times0.51(177)$&   $0.38\times0.31(161)$     & 103 (2.9)    & 15             & 171.1 (4.7)  \\ 
347.9       &   354.2     &       57.293$^s$, 11.38$''$              &   $0.58\times0.51(178)$&   $0.36\times0.32(155)$     & 115 (3.1)    & 15             & 191.6 (5.1)  \\ 
\hline
\end{tabular}
\tablefoot{Results from the double Gaussian profile fits to the continuum emission in the image plane.
\tft{a}{Peak brightness temperatures as described in \citet{Wilson2014} and taking 81, 77, 58, and 56~K/Jy conversion factors for 247.0, 263.1, 336.9, and 347.9~GHz, respectively.}
}
\end{table*}

\subsection{HCN and HCO$^+$}
\label{Sect.HCNHCOp}

\begin{figure*}
\centering
\includegraphics[width=\linewidth]{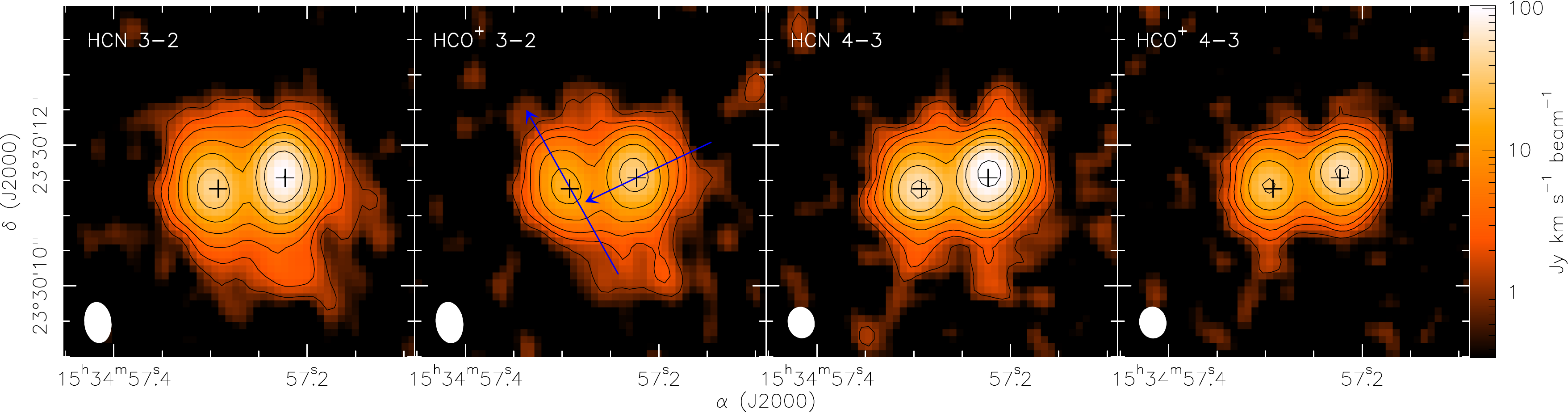}
\caption{Integrated intensities of HCN $3-2$, HCO$^+~3-2$, HCN $4-3$, and HCO$^+~4-3$ between $-400$ and 400~\kms
around the systemic velocity (see Sect.~\ref{Sect.HCNHCOp}). Contours represent 3, 6, 10, 20, 40, 80, 150, and 300~$\sigma$ levels where $1\sigma=0.35~\rm Jy~km~s^{-1}~beam^{-1}$. Peaks of continuum emission (Fig.~\ref{fig.continuum}) are indicated with crosses.
The region shown is the same as in Fig.~\ref{fig.continuum}.
The blue arrows in the HCO$^+~3-2$ panel indicate the direction ($P.A.=30^\circ$ and $115^\circ$) of the P-V cuts in Fig.~\ref{fig.pvdiagrams}.
 \label{fig.integrated}}
\end{figure*}

Despite the short snapshot observations, line emission of HCN and HCO$^+$ in the $J=3-2$ and $J=4-3$ transitions is detected at unprecedentedly high signal-to-noise ratio of up to $\sim300$.

In Fig.~\ref{fig.integrated} we show the intensities integrated over the central $800$~\kms~ of each transition.
Similar to what is observed in the integrated intensity of CO $6-5$ at high resolution \citep{Rangwala2015}, 
we observe an offset between the peak of molecular emission and the continuum, which is more evident toward the WN in both the HCN and HCO$^+$ $J=4-3$ transitions.
The measured differences of $\sim0.04''$ are most likely due to the effect of the absorption systems (Sect.~\ref{sect.abssystems})
in their integrated CO map.
Indeed, \citet{Rangwala2015} concluded from their kinematic models that this offset in CO is only apparent. 

\begin{figure}
\centering
\includegraphics[width=\linewidth]{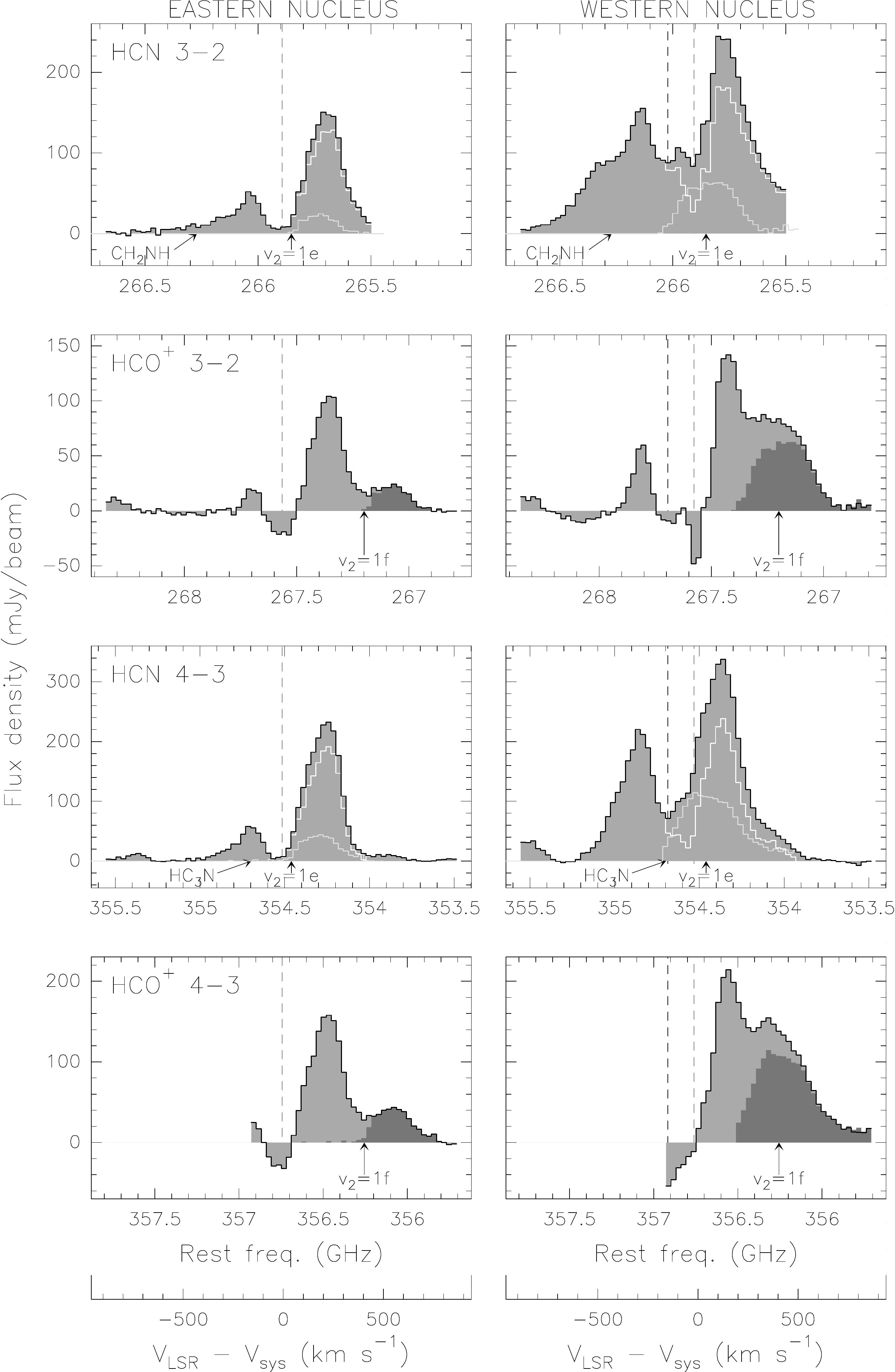}
\caption{
Extracted spectra from the two nuclei in Arp~220 shown as gray histogram and black continuous lines.
X-axes are labeled both in rest frequency and at a velocity offset relative to the systemic value (see Sect.~\ref{Sect.HCNHCOp}).
Observed profiles from the eastern and western nucleus are shown in the left and right columns, respectively.
The transition that mostly contributes to the observed spectra is displayed in the top left corner of the left column panels.
Vertical dashed lines correspond to the apparent velocities of the absorption systems that affect the line profile for each nucleus (Sect.~\ref{sect.abssystems}).
The expected position at the systemic velocity of the HCN~$v=1e$ and $1f$ lines are indicated with labeled arrows.
The dark histograms show the  HCN~$v=1f$ residual emission after subtracting an estimated line profile (based on HCN) from the HCO$^+$ line (see Sect.~\ref{sect.vibresults}). 
Pale gray lines show the possible contribution of HCN~$v=1e$ to the HCN emission, while the white lines represent the HCN emission after subtracting
the contribution from HCN~$v=1e$ (Sect~\ref{sect.vibContrib}).
The position of the potential lines of CH$_2$NH and HC$_3$N blended to HCN $3-2$ and $4-3$, respectively, are also indicated.
\label{fig.lineprofiles}}
\end{figure}

Figure~\ref{fig.lineprofiles} shows the spectral profiles of the four transitions extracted from the positions of the two nuclei defined as the peak of continuum emission in Sect.~\ref{sect.continuum}.
The velocities we report here are referenced to $z=0.018126,$ which correspond to $V_{\rm LSR}=5434$~\kms~ (optical) or 5337~\kms~(radio).
We refer to this as the systemic velocity ($V_{\rm sys}$) of Arp~220 and use the radio convention of velocities throughout the paper.
The individual systemic velocities of each nucleus are denoted as $V_{EN}$ and $V_{WN}$.

Figure~\ref{fig.integratedranges} shows integrated maps for each of the four transitions in ranges of 200~\kms~ between $-600$ and 600~\kms.
Position velocity diagrams in the direction of the velocity gradient of each nucleus ($P.A.=30^\circ$ and $115^\circ$ for EN and WN, respectively) are presented in Fig~\ref{fig.pvdiagrams}.  The directions of the cuts are also indicated by blue arrows in Fig.~\ref{fig.integrated}. 

\subsubsection{Line blending}
Line blending in Arp~220 is severe and cannot be neglected \citep{Mart'in2011}. While emission from vibrationally
excited HCN is discussed in Sect.~\ref{sect.vibresults}, we here
discus the possible contamination from other lines close to the HCN and HCO$^+$ lines.

From the Arp~220 and NGC~4418 line surveys \citep{Mart'in2011,Costagliola2015} we know that the compact obscured nuclei in these sources present large abundances of HC$_3$N, which contribute to their spectra with transitions
every $\sim9$~GHz, together with a comb of vibrationally excited transitions.
In the frequency range studied here, we find the HC$_3$N $38-37$ transition at 354.7~GHz.
From the physical conditions derived by \citet{Mart'in2011}, we infer for this line a peak flux density integrated over the whole
source of $\sim20-40$~mJy within the uncertainties. We can therefore roughly estimate peak flux densities of $\sim10-20$ mJy if they
are equally divided between the nuclei. Thus it will slightly contribute to the blueshifted side of HCN $4-3$ of the EN, and right at one of the absorption systems (Sect.~\ref{sect.abssystems}) in the WN.
This contribution will be even stronger if a warm HC$_3$N component ($T_{\rm ex}>40$~K) is present, as seems to exist in the ALMA data for NGC 4418 \citep{Costagliola2015}.
Contribution from vibrationally excited transitions of HC$_3$N are expected to be about half of the peak flux of the ground-state transitions \citep{Mart'in2011}.
We note that, as discussed in Sect.~\ref{sect.vibContrib}, and depending on the relative spatial distribution, the emission from this species might be fully absorbed by foreground optically thick HCN and thus does not contribute to the observed HCN profile.

The bump observed at the blueshifted side of HCN $3-2$ toward the WN and less prominently toward the EN might be due to the CH$_2$NH $4_{1,3}-3_{1,2}$ transition at 266.3~GHz. This feature is better identified in the narrower spectral lines
observed in the LIRGs IC860 and Zw49-57 presented by \citet{Aalto2015a}. However, this has to be confirmed with other transitions of this molecule for a proper identification in Arp~220.

\subsubsection{Absorption systems}
\label{sect.abssystems}
We observe clear absorption features that have previously been identified in every high-resolution imaging of
CO \citep{Downes2007,Sakamoto2008,Rangwala2015}, HCO$^+$ \citep{Sakamoto2009},
HCN, CS, CN \citep{Scoville2015}, and even SiO \citep{Tunnard2015}.
The apparent absorption peaks are highlighted with vertical dashed lines in Fig.~\ref{fig.lineprofiles}.
However, to infer the actual central velocity of the absorption features, we also have to account for the contribution from the emission profile.

To characterize the absorption profiles, we use the HCO$^+$ $3-2$ transition, which is the least affected by blending and also fully covered by the
bandwidth of our observations. We assumed a Gaussian shape for the HCO$^+$ emission and performed a fit to the unabsorbed channels, while masking those where the absorption is observed. After subtracting the emission profile, we fit the individual absorptions.
We note that these numbers are highly dependent on the line profile of the emission, but we considered the Gaussian as the best approach at the current resolution. 
For this reason, we do not refer to the statistical errors of the fits since they are not meaningful.
In Fig.~\ref{fig.specabs} we show the multi-Gaussian fit performed to the HCO$^+$ $3-2$ profiles that are discussed in the following.

Toward the EN, the central velocity of the absorption appears at $V-V_{sys}=-5\pm4$~\kms~  and a line width of $\sim250\pm10$~\kms.
The HCO$^+$ $3-2$ Gaussian fit to the unabsorbed channels yields a central velocity of $\sim117\pm8$~\kms~ and width $\sim290\pm10$\kms.
Because of this line emission, the absorption profile is shifted to $\sim70$~\kms~ with a narrower width of $\sim210$\kms.
Thus only one absorption component is observed in the EN at $V\sim5410$\kms.

Toward the more complex WN, the apparent velocities of the absorptions are at $V-V_{sys}=-150\pm8$ and $20\pm5$~\kms~ with widths of $210\pm30$ and $126\pm9$~\kms, respectively.
The Gaussian fit to the HCO$^+$ $3-2$ emission is centered on $\sim5\pm4$~\kms~ with a width of $\sim386\pm10$~\kms.
After subtracting this profile, the absorption lines appear to be centered on $\sim-135$ and $11$~\kms~ with widths of
210 and 130~\kms, respectively.
Our corrected blueshifted absorption agrees better with the average velocity centroid of $-110\pm10$~\kms~(relative to $z=0.018126$) measured with several HCN IR absorption profiles \citep{Gonzalez-Alfonso2013}, which supports our result and thus the assumption of an underlying Gaussian emission profile.

Additionally, we detect a broad absorption profile centered on $V-V_{sys}=-570\pm20$\kms~ and with a width of $\sim170\pm40$\kms\ only toward the WN.
This feature was previously observed with the SMA also in HCO$^+$ $3-2$ \citep{Sakamoto2009} and more recently in SiO \citep{Tunnard2015}.
This means that a total of three absorption systems are identified at $V_{\rm LSR,radio}\sim4767$, 5202, and 5348~\kms. 
Although they are present in the SMA data from \citet{Sakamoto2009}, the absorption of the three profiles shows a similar depth in their data, while in the ALMA data the absorption close to the systemic velocity is significantly deeper than the other two. Moreover, while the disagreement between the two central absorptions
can be attributed to the different quality of the data and the fitting to the profiles, the extreme blueshifted absorption reported at 4890~\kms~ by \citet{Sakamoto2009} is $\sim120$~\kms~closer to the Arp~220 systemic velocity than in our ALMA data.

We did not detect any trace of redshifted absorption in either HCN or HCO$^+$ such as that reported for CO $6-5$ by \citet{Rangwala2015}.
Moreover, we did not observe the bow shape in the integrated emission as seen in CO $6-5$.
Judging from the differences in the channel maps of CO and those of HCN and HCO$^+$, the attenuation of the emission toward the SW side in the two nuclei could be due to the redshifted material of lower density that is observed in absorption in CO $6-5$
\citep[$n_{crit}\sim10^5~cm^{-3}$,][]{Rangwala2015} but not in the high-density tracers HCN and HCO$^+$ \citep[$n_{crit}\sim10^6$ and $10^7~cm^{-3}$, respectively,][]{Shirley2015}. 

\subsubsection{Individual velocities of the nuclei ($V_{WN}$ and $V_{EN}$)}
\label{sect.vsys}
Precise estimates of the systemic velocities of the individual nuclei are important parameters to determine whether there are radial motions to or from the nuclei.
Previous estimates based on the CO profiles resulted in radio velocities of $V_{WN}=5355\pm15$~\kms~ and $V_{EN}=5415\pm15$~\kms~ \citep{Sakamoto2009}.

High-density tracers probing an inner region can better pinpoint the systemic velocity of the individual nuclei.
The fit to the unabsorbed parts of the HCO$^+$~$3-2$ profiles described in Sect.~\ref{sect.abssystems} and shown in Fig.~\ref{fig.specabs} can be used as a good approximation.
The velocity offsets of $V-V_{sys}\sim5$ and $\sim117$~\kms~ found for the WN and EN, respectively, result in individual nuclear velocities of $V_{WN}=5342\pm4$~\kms~ and $V_{EN}=5454\pm8$~\kms.
Based on our data, the velocity difference between the two nuclei is then $112\pm9$~\kms.
This result is consistent within the uncertainties with that of \citet{Sakamoto2009} toward the WN, while the EN velocity is $\sim40$~\kms~ above their estimate based on CO.

A precise knowledge of the systemic velocity is a key parameter in any attempt of modeling the kinematics of the Arp~220 nuclei. The CO $6-5$ kinematics in the EN were modeled with an optically thick turbulent rotating disk by \citet{Rangwala2015}. In such a self-absorbing disk model, absorption is expected to be centered on the systemic velocity of the EN , which \citet{Rangwala2015} estimated at 5345~\kms.
However, we showed that the absorption in the EN occurs at $v\sim5410$~\kms~ (Sect.~\ref{sect.abssystems}), which is blueshifted by 50~\kms~from our derived $V_{EN}$.
Toward the WN, we found one of the three absorption systems centered directly on the derived nuclear velocity with $V-V_{WN}\sim6$~\kms~ (with the other two blueshifted to  $V-V_{WN}=-140$ and $-575$~\kms).
For the absorption systems in other compact and highly obscured galaxies, \citet{Aalto2015a} assumed the absorption to be centered on the systemic velocity, as expected from a uniformly distributed cold gas envelope. However, a close inspection of the Arp~220 system may show that the absorption is produced by outflowing material close to the nuclear region (see Sect.~\ref{sect.absimaging}).

The full spectral scan (Mart\'in et al. in preparation) is expected
to provide further constraints on the systemic velocity of the two nuclei, with a large number of optically thin lines with high critical densities less affected by absorption effects.

\subsection{HCN~$v_2=1$ emission}
\label{sect.vibresults}
So far, only \citet{Sakamoto2010} have presented the detection of two transitions of HCN~$v_2=1$ toward NGC 4418.
We detected the emission of HCN~$J=3-2$ (267.199~GHz) and $4-3$ (356.255) in its $v_2=1f$ vibrational state. 
The latter has been reported toward the WN by \citet{Aalto2015a}. Here we report the emission clearly detected toward the two nuclei in both transitions.
Although we do not report it there, HCN~$v_2=1f$ emission can also be identified in the SMA data from \citet{Sakamoto2009},
whose sensitivity is lower.

The HCN~$v_2=1f$ transitions are separated by $\sim402$~\kms~ from the HCO$^+$ lines. Thus the vibrationally excited line is significantly blended with HCO$^+$ toward the WN and to a lesser degree with the redshifted HCO$^+$ emission in the EN (Fig.~\ref{fig.lineprofiles}).
 To fit the emission of HCN~$v_2=1f$ in both nuclei, we followed two simple approaches. The results are presented
in Table~\ref{tab.vibfit}.
First we  fit the redshifted peak of HCO$^+$ and the HCN~$v_2=1f$ line with a double Gaussian. This is indicated in Table~\ref{tab.vibfit} as {\it double}. However, this fit does not take into account the tail of redshifted emission observed in both HCN $3-2$ and $4-3,$ which might be also present in HCO$^+$.
This is particularly critical in the WN. 
To account for a more realistic HCO$^+$ line shape, we subtracted a scaled-down profile from HCN $3-2$ to subtract
the HCO$^+$ emission and then fit the HCN~$v_2=1f$ emission to the residual. 
The possible contribution from $\nu_2=1e$ (Sect.~\ref{sect.vibContrib}) was not removed from the subtracted HCN $3-2$ profile.
The residuals after subtraction of the assumed HCO$^+$ profile are shown in Fig~\ref{fig.lineprofiles} as
dark gray histograms. Results of the Gaussian fit to these residuals appear in Table~\ref{tab.vibfit} as {\it subtracted}.
Although subtracting the HCN $3-2$ profile to account for HCO$^+$ is highly uncertain, the central velocities derived from the two HCN~$v_2=1f$ transitions show agree better than when derived from a double Gaussian fit.
Only higher spatial resolution observations will be able to accurately determine the line profile
of the vibrational emission of HCN.

If our de-blending method is correct, the vibrational emission appears to be redshifted by $15\pm11$ and $22\pm9$~\kms~ with respect to the derived nuclear velocities $V_{WN}$ and $V_{EN}$ (Sect.~\ref{sect.vsys}), respectively. At our spectral resolution and fit accuracy, this is consistent with this emission being centered on the individual nuclear velocities. These velocities
support the scenario that vibrational emission traces the innermost regions around the nuclei (Sect.~\ref{sect.vibHCN}).

\subsubsection{Contribution of $v_2=1e$ to the HCN $v=0$ profile}
\label{sect.vibContrib}
Based on its spectroscopic parameters, the emission of the $v_2=1e$ transitions at 265.853~GHz ($J=3-2$) and 354.460~GHz ($J=4-3$) are expected to have similar flux densities as those of the $v_2=1f$ lines.
These $v_2=1e$ transitions appear at $\sim+38$~\kms~ from the center of the HCN ground-vibrational transitions, thus possibly affecting its profile.
Assuming a similar emission to the $v_2=1f$ fitted profiles, we have estimated the contribution of this emission to the $v=0$ profiles at the two nuclei.
In Fig.~\ref{fig.lineprofiles} the estimated $v_2=1e$ profiles are shown as pale gray lines.
The residual HCN $v=0$ after subtracting the estimated contribution from the $v_2=1e$ is shown as white lines.

However, we note that the profiles presented in Fig.~\ref{fig.lineprofiles} correspond to the limit case in which the $v_2=1e$ contribution is not attenuated by foreground material. 
Indeed, the $v_2=1e$ line may be partially or totally absorbed by the foreground HCN ground transition. In this case, no contribution to the observed profile would be observed.
Vibrational emission might escape if HCN were absorbing gas that
is clumpy or not equally spatially distributed per velocity bin, although the velocity offset between the lines might not be enough for the latter possibility.
Because of this uncertainty, we did not take the $v_2=1e$ contribution into account to derive the HCN line profile that was used to de-blend HCO$^+$ from the $v_2=1f$ lines (Sect.~\ref{sect.vibresults})

\begin{table*}
\caption{Gaussian fits to the HCN~$v_2=1f$ emission in the two nuclei \label{tab.vibfit}}
\begin{tabular}{l c c c c c c c c c}
\hline
Line                   &  Rest Freq.   &   Fit\tfm{a}   & $v_0-v_{sys}$\tfm{b} &  $v_0-v_{N}$\tfm{c}   &   $\Delta v_{1/2}$    &   $S_0$    &  $\int S \delta v$   &  $L_{vib}$         &  $\frac{L_{vib}}{L_{FIR}}$\tfm{d} \\
                       &    (GHz)      &                &   \kms               &    \kms               &     \kms              &     mJy    &     Jy~\kms          &  $10^{3}~L_\odot$  &  $10^{-8}$                  \\
\hline
\hline
                    \mc{10}{c}{\large \bf Western Nucleus}  \\
HCN $J=3-2,v_2=1f$     &    267.199    &    $double$    &   -33 (4)            & -38                   &    354 (9)            &    90      &    33.8 (0.8)        &  58                &      9.2              \\
                       &               &    $subtract$  &   24 (4)             &  19                   &    296 (10)           &    67      &    21.2 (0.6)        &  36                &      5.7              \\
HCN $J=4-3,v_2=1f$     &    356.255    &    $double$    &   -57 (5)            & -63                   &    424 (10)           &    148     &    67.4 (1.5)        & 154                &     24.4              \\
                       &               &    $subtract$  &   14 (11)            &  9                    &    353 (33)           &    112     &    42 (3)            &  96                &     15.2              \\
                     \mc{10}{c}{\large \bf Eastern Nucleus}  \\
HCN $J=3-2,v_2=1f$     &    267.199    &    $double$    &   120 (10)           &  3                    &    220 (30)           &    23      &    5.6 (0.5)         &  10                &     20.0              \\
                       &               &    $subtract$  &   140 (2)            &  23                   &    179 (5)            &    24      &    4.6 (0.1)         &   8                &     16.0              \\
HCN $J=4-3,v_2=1f$     &    356.255    &    $double$    &   132 (2)            &  15                   &    242 (4)            &    43      &    11.2 (0.2)        &  26                &     51.9              \\
                       &               &    $subtract$  &   143 (2)            &  26                   &    209 (5)            &    45      &    10.0 (0.1)        &  23                &     45.9              \\
\hline
\end{tabular}
\tablefoot{Results from the Gaussian fit to the HCN $v_2=1f$ transitions.
\tft{a}{Fit procedure: ($double$) Simultaneous double Gaussian fit to high-velocity HCO$^+$ $v=0$ and HCN~$v_2=1f$ features; ($subtract$) Fit to $v_2=1f$ after HCN emission scaled down to the HCO$^+$ intensity is subtracted (see Sect.~\ref{sect.vibresults} for details.)}
\tft{b}{Central velocity of the Gaussian profile with respect to the systemic velocity of the entire system.
\tft{c}{Velocity with respect to the individual velocities, $V_{EN}$ and $V_{WN}$, derived for each nucleus (Sect.~\ref{sect.vsys})}
\tft{d}{Assuming $log(L)=11.8$ and $10.7~L_\odot$ for the western and eastern nuclei, respectively \citep{Wilson2014}}}
}
\end{table*}



\subsubsection{Gas surface densities in the Arp~220 nuclei}
\label{sect.mvir}

As discussed in Sect.~\ref{sect.sizepositionvelocity}, vibrational emission from HCN is emerging from the very central $50-60$~pc regions around the two nuclei in Arp~220.
This makes it an excellent probe of the inner enclosed mass within the central few tens of parsec.

The average FWHM derived from the two observed
transitions are $325\pm25$~\kms~ and $195\pm15$~\kms~ for the WN and EN, respectively.
If all the vibrationally excited emission arises from within $r\sim25-30$~pc (Sect.~\ref{sect.sizepositionvelocity}), we can estimate virial masses as
 $(M_{vir}/M_\odot)=250(\Delta v_{1/2}/{\rm km~s^{-1}})^2(R/{\rm pc})$ \citep{Wilson2009} of 
$M_{WN}\sim6-9\times10^8$ and 
$M_{EN}\sim2-3\times10^8~M_\odot$.
These values are an order of magnitude lower than the estimates of $\sim6\times10^9~M_\odot$ within the central
$\sim100$~pc of the two nuclei \citep{Engel2011}. We note that if we were to calculate the virial masses as $M_{vir}=3.36\sigma^2 R/G$ \citet{Engel2011}, our derived masses would be a factor of $\sim2$ lower.
%

These masses correspond to gas surface densities of $3\pm0.3\times10^4~M_\odot~\rm pc^{-2}$ (WN) and $1.1\pm0.1\times10^4~M_\odot~\rm pc^{-2}$ (EN) assuming a gas fraction $f_g=0.1$. These values are roughly one order of magnitude lower than
the values derived by \citet{Barcos-Munoz2015}.
Still, both nuclei may be close to the maximal stellar surface density of $\sim 10^5~M_\odot~\rm pc^{-2}$ \citep{Hopkins2010} for plausible higher gas fractions.

\section{Dense gas absorption}

\begin{figure*}
\centering
\includegraphics[width=\linewidth]{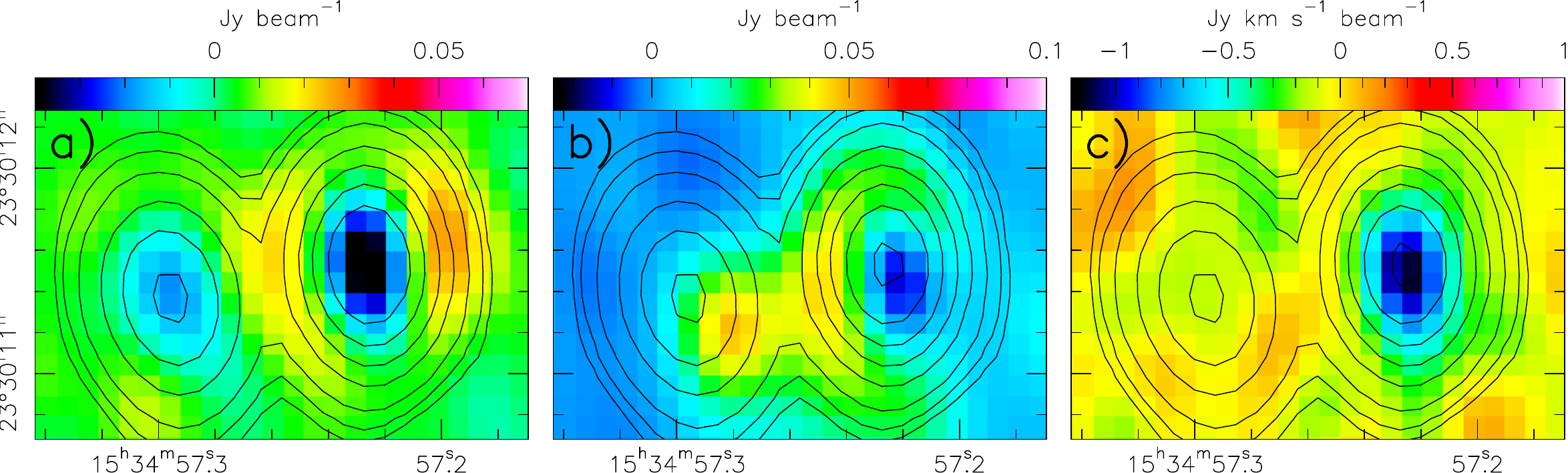}
\caption{In color we represent the different absorption systems identified
in the nuclei of Arp~220 as observed in HCO$^+$ $3-2$. Overlaid, the contours represent the continuum emission. Panels $a)$ and $b)$ show 25~\kms~ individual channels centered
on 0 and $-150$~\kms, respectively. Panel $c)$ presents the integrated emission in the velocity range $[-625, -525]$~\kms.
\label{fig.absorptions}}
\end{figure*}

\subsection{Underestimated dense gas in compact obscured nuclei}
As has been suggested from the CO $6-5$ observations from \citet{Rangwala2015}, absorption in Arp~220 severely affects the dense gas observations.
This is fully supported by our observations of HCN and HCO$^+$.
However, we find that the peak of molecular emission in the HCN ground state matches the 
continuum peak. Both the highest HCN integrated intensity and the peak intensity are found at the continuum peak position.
These observations further prove that the offset between the continuum peak and the molecular peak seen in CO $J=2-1$ \citep{Downes2007}, CO $J=3-2$ \citep{Sakamoto2008,Sakamoto2009}, and CO $J=6-5$ \citet{Rangwala2015} is due to the stronger effect of the absorption in CO than in HCN and HCO$^+$.

If, as assumed in Sect.~\ref{sect.abssystems}, we consider the real line profiles of the ground-state transitions to be Gaussian in shape, we estimate that on average $\sim 70\%$ of the line profiles in the two Arp~220 nuclei could be absorbed. 
Under the assumption of a 3D Gaussian distribution of the emission,
where the extent of the line emission would be equal to that of the absorption (which is not true, as seen in Sect.~\ref{sect.absimaging}), these $70\%$ would be an absolute upper limit to the absorbed total integrated flux density.
Although this absorption is concentrated in the very central region, low-resolution observations
over $\sim$kpc scales still show these absorbed profiles \citep{Greve2009,Mart'in2011,Aladro2015}.

Absorption in dust-obscured nuclei such as those of Arp~220 are clearly not an isolated case and might be systematically affecting the most compact LIRGs and ULIRGs \citep{Aalto2015a}.
Absorbed profiles might result in restrictions to the use of HCN and
HCO$^+$ as probes of the overall dense gas content in this type of galaxies and more
importantly restrict estimates of their star-forming efficiencies (SFE) \citep{Gracia-Carpio2008}.
\citet{Gracia-Carpio2008} showed evidence of an enhancement of the dense gas SFE, measured as the $L_{\rm FIR}/L'_{\rm HCN(1-0)}$ ratio in a sample of LIRGs and ULIRGs with respect to normal galaxies.
Spectra in their sample also showed a variety of line shapes with flat tops and dips in the profile that were commonly attributed to gas kinematics.
If the line profiles from \citet{Gracia-Carpio2008} are the result of cool gas foreground absorption, their total amount of dense gas would be underestimated by more than a factor of 2 in these galaxies, and therefore the SFE would be overestimated.
Although this effect would only be expected in the most compact extreme objects, this uncertainty could result
in a strong bias, and caution should be applied on the higher end of galaxy luminosities.

\subsection{Origin and location of the absorption systems}
\label{sect.absimaging}
Whether the absorption is due to self-absorption and/or absorption of the continuum cannot be distinguished at our $0.5''$ resolution
since both the continuum and the hot dense gas material that
is traced appear to be centered on the same position.
Self-absorption would require a strong temperature gradient, which is present, as we can deduce from the detection of vibrationally excited emission of HCN (Sect.~\ref{sect.sizepositionvelocity}). However, the detection of HCO$^+$ absorption below the continuum level shows that a profile like this is at least partly due to absorption of the continuum.

To understand the origin of the absorption systems, in Fig.~\ref{fig.absorptions} we image the absorption systems identified in Sect.~\ref{sect.abssystems}.
Images are centered on the apparent velocities of the absorptions and not on the corrected velocities
(see Sect.~\ref{sect.abssystems}), where the absorption is contaminated by emission.
We used the HCO$^+$ $3-2$ transition because it shows absorption of the continuum, and therefore the location
of the absorption is more easily visualized.
From Fig.~\ref{fig.absorptions} we observe that the absorption found closer to $V_{sys}$, toward both the EN and WN, is located
directly on top of the peak of the continuum emission.

On the other hand, the absorption observed at $V-V_{sys}\sim -150$~\kms~ is marginally offset by $\sim0.05''$ (20~pc) southwest from the nuclear position.
This absorbing blueshifted gas most likely traces gas outflowing from the WN.

Similarly, the absorption system at $V-V_{sys}\sim-570$\kms~also appears to be centered on the peak of continuum emission or just marginally south $\lesssim0.05''$ of this position.

Toward the WN, \citet{Tunnard2015} found an offset of $0.1"$ between the location of the SiO $J=6-5$ absorption and the emission in the WN close to systemic velocity.
This offset was interpreted as a bipolar outflow, although it may be partly caused by the location of the absorbing system. 
From our HCN and HCO$^+$ observations, the absorption at $V_{sys}$ is dominated by the gas along the line of sight with no sign of an offset between the absorption and the emission. The location of the emission of HCN and HCO$^+$ (Fig.~\ref{fig.absorptions}$a$)
differs from what  \citet{Tunnard2015} observed in SiO. This supports their idea that SiO is emitted in the gas affected by the putative outflow, but higher resolution imaging is required
to conclude about this question.

The absorption observed in HCN and HCO$^+$ at $V-V_{WN}\sim6$~\kms~ is very likely related to the cooler
structure in the outer regions of the WN that both self-absorbes the molecular emission and absorbes
the extended background continuum emission, but cannot be directly associated with outflowing gas.
However, the blueshifted absorptions at $-140$ and $\sim-575$~\kms~ relative to $V_{WN}$ are associated with high-velocity gas outflowing from the two nuclei.
Their offsets with respect to the peak of the continuum emission agree with the sketch proposed by \citet{Tunnard2015}, perpendicular
to the plane of rotation of the molecular disk.

For the high velocity $\sim-570$\kms~component, the redshifted counterpart of the outflow might be responsible for the line wing we observe in both HCN $3-2$ (cropped at the edge of the spectrum) and $4-3$ toward the WN. As shown in Fig.~\ref{fig.integratedranges}, HCN emission in the $400-600$~\kms~ interval
is offset toward the north, as expected for a bipolar outflow, and contrary to what is observed in HCO$^+$
, where the emission is dominated by the vibrationally excited HCN emission.
However, the line confusion affecting this galaxy prevents us from firmly confirming this redshifted outflow component at the current resolution.
Thus, the absorption observed in SiO close to systemic velocity \citep{Tunnard2015} and the other absorption systems may all be related to different components of the same outflow aligned in the northeast to southwest direction.

Toward the EN the absorption appears to be centered on the peak of continuum emission, even though the absorption is blueshifted by $-50$~\kms~ relative to $V_{EN}$, which also suggests outflowing material from this nucleus.

\section{Vibrationally excited HCN}
\subsection{HCN excitation of vibrational levels}
\label{sect.vibHCN}
Based on the discussion on recent detections of vibrationally excited HCN in extragalactic objects, collisional excitation of HCN up to
the vibrational levels has been discarded \citep{Sakamoto2010,Imanishi2013,Aalto2015a}.
Vibrational HCN levels are efficiently excited by 14$\mu$m continuum photons provided that $T_{\rm B}>100$K \citep{Aalto2015a}.
Our data show that the peak of the continuum emission matches the position at which vibrationally excited emission is found, which supports the idea that the emission is excited through continuum photons.
Far-IR observations measure the photospheric $T_{dust}=90-140$~K from line excitations of various species such as H$_2$O or OH \citep{Gonzalez-Alfonso2012}. Since the inner $T_{dust}$ will be higher, it would be compatible with 14$\mu$m photon pumping.

If we assume local thermodynamic equilibrium (LTE) conditions, the excitation temperature of the HCN based on the integrated intensities of the $v_2=1f$ $J=3-2$ and $4-3$ transitions is $T_{\rm ex}\sim38$~K and $45$~K for the WN and EN, respectively. Statistical errors in the fit to the transitions are irrelevant since the main source of uncertainty lies in the blending of the $v_2=1f$ transitions.
If we do not invoke radiative pumping in the WN for such an excitation temperature, a column density of $N_{HCN}\sim 10^{24}~\rm cm^{-2}$ would be required to account for the observed line intensities,
together with a $0.5''$ source size. This column density would reach a few $10^{25}~\rm cm^{-2}$ for smaller source sizes of $\sim0.2''$.
 Although numerically
feasible if the medium is dense enough \citep[$n_{crit}>10^{11}~\rm cm^{-3}$][]{Ziurys1986}, it would therefore impose unrealistic constraints on the HCN abundances.
 
We further note that a hot interstellar medium ($T_{\rm}\gg200$~K) could explain the observed flux densities for  $N_{HCN}\ll10^{18}~\rm cm^{-2}$ (for a $0.5''$ source). However, under these conditions, the optically thin ratio of the two $v_2=1f$ transitions should be $J=4-3/3-2\sim3$, which does not match the observed ratio of $\sim2$.
These basic arguments, which also apply to the EN, further confirm the impossibility of understanding HCN excitation without taking into account radiative
excitation through IR photons.

The derived excitation temperature of $\sim 40$~K is consistent with what is observed in NGC~4418 \citep{Sakamoto2010}, which, apart from Arp~220, is the only source for which two HCN transitions in the $v_2=1$ vibrational level have been reported.
We cannot reliably estimate the vibrational temperature as measured between the $v=0$ and $v_2=1f$ transitions because of the heavy
line absorption and blending features that affect HCN. In addition, HCN is probably optically thick.
Observations of HC$_3$N in the vibrationally excited levels probe $T_{\rm vib}=350$~K between vibrational levels with 
$T_{\rm ex}=35$~K measured within the ground-vibrational state \citep{Mart'in2011}.
This is again quite similar to the excitation conditions measured in NGC~4418 \citep{Costagliola2010,Sakamoto2010}, as has been
noted by \citet{Costagliola2015}.
A comprehensive explanation on the excitation of HCN vibrational levels was presented by \citet{Sakamoto2010} for NGC~4418, which also applies to the conditions we observe
in Arp~220.

By only visually inspecting the spectra in Fig.~\ref{fig.lineprofiles}, we see that the ratio between the peak flux density of the HCN $v=0$ to the $v_2=1f$ line seems to be higher in the EN. This would imply a lower $T_{\rm vib}$ , or possibly a weaker pumping effect on the excitation of HCN toward this nuclei.
However, the HCN line profiles again provide little information because of blending, absorptions, and optical thickness. From the isotopic ratio observations
of H$^{13}$CN $3-2$ by \citet{Tunnard2015}, we infer a ratio between the flux densities of the two nuclei of WN/EN$=2-3,$ depending on whether we consider the integrated or peak flux densities. Although it is also affected by absorption, this ratio is probably less affected by the opacity effects.
A proper temperature estimate would require observations of the two lines at similar resolution, given the different extent of the H$^{13}$CN $v=0$ and the HCN $v_2=1$ states.
Qualitatively, the observed ratio between the nuclei in the $v_2=1f$ transition is W/E$\sim2,$ which implies that the $T_{\rm vib}$ in the EN is probably similar to that in the WN.
Further conclusions on the vibrational temperature will be derived from the observations at similar resolution of the HCN isotopologs with ALMA (Mart\'in et al. in preparation) and from the spatially resolved HC$_3$N vibrational emission.

The excitation arguments above assume the vibrationally excited levels to be optically thin. However,  these transitions could also be optically thick (Gonz\'alez-Alfonso in preparation).
In the optically thick regime, the expected flux density line ratio is $J=4-3_{v2=1f}/3-2_{v2=1f}\sim1.75$. Based on the peak flux densities, the observed ratios are 1.7 and 1.9 for the WN and EN, respectively (2.0 and 2.2 based on the integrated flux densities).
Thus, the $v_2=1f$ emission might be optically thick. This would imply that the emission originates from a very compact area. 
Taking as limiting cases the excitation temperature measured above $T_{ex}\sim40$~K  and that derived from HC$_3$N vibrational levels of $T_{vib}\sim300$~K, the observed HCN $V_2=1f$ peak flux densities in the WN would imply
sizes of $0.18''$ (66~pc) and $0.06'' $ (22~pc), respectively,
while toward the EN, the observed peak flux densities would yield sizes of $0.11''$ (40~pc) and $0.04''$ (15~pc).
These sizes derived in the optically thick regime are lower limits to the actual size of the emitting regions toward the two nuclei.

\subsection{Vibrational emission within the opaque dust}
\label{sect.sizepositionvelocity}

\begin{figure*}
\centering
\includegraphics[width=\linewidth]{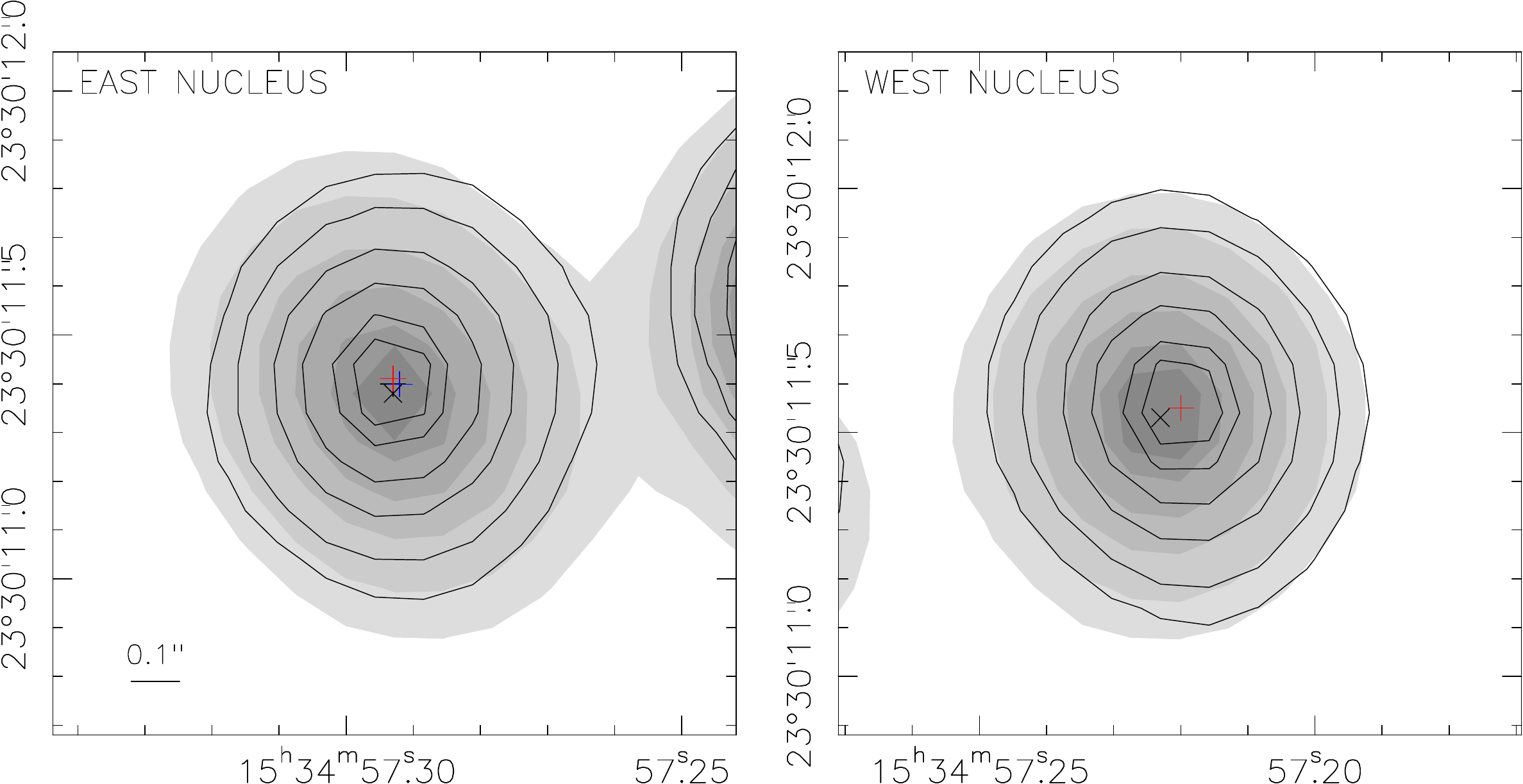}
\caption{Integrated flux density of the vibrationally excited HCN $4-3$ (in contours) superimposed on the 347.9~GHz continuum emission (grayscale). Integrated velocities are 40 to 240~\kms~ for the eastern nucleus, covering the whole line, and from  20 to 170~\kms (the redshifted region is not affected by blending with HCO$^+$) for the western nucleus.
Contours are 10, 20, 40, 60, 80,and 90\% of the peak-integrated flux density.
The black cross represents the fitted center to the continuum emission, while the black, red, and blue plus signs represent the fitted centers of the vibrational line for the integrated, redshifted, and blueshifted parts of the line profile (see text in Sect.~\ref{sect.sizepositionvelocity}), respectively. For the western nucleus only the redshifted emission was available because of blending with HCO$^+$ on the blueshifted side.
\label{fig.vibmaps}}
\end{figure*}

\begin{table*}
\caption{Gaussian fit to $J=4-3~v_2=1f$ maps \label{tab.vibimagefit}}
\begin{tabular}{l c c c c c c }
\hline
Vel.range~\tfm{a}   &     offset~\tfm{b}        &  Convolved Size           &   Deconv. Size             &   Peak Int. Flux          & Int. Flux      & \\
(\kms)              &     ($'',''$)             &($''\times''~(^\circ)$)   & ($''\times''~(^\circ)$)   &   (Jy~\kms~beam$^{-1}$)   &   (Jy~\kms)   &\\
\hline
\hline
                    &        \mc{5}{c}{\large \bf Western Nucleus}  \\
20-170              &   0.04,0.02               &  $0.47\times0.41~(12)$    &  $0.16\times0.12~(50)$     &  18.5 (0.6)               &  20.6 (1.1)    &\\
                    &        \mc{5}{c}{\large \bf Eastern Nucleus}  \\
40-240              &   0.00,0.02               &  $0.47\times0.31~(179)$   &  $0.17\times0.12~(120)$    &   8.3 (0.1)               &   9.3 (0.3)    &\\
40-140              &   0.00,0.02               &  $0.47\times0.42~(175)$   &  $0.17\times0.08~(7)$      &   5.0 (0.1)               &   5.6 (0.2)    &\\
140-240             &   0.01,0.03               &  $0.47\times0.42~(1)$     &  $0.16\times0.11~(120)$    &   5.0 (0.1)               &   5.6 (0.2)    &\\ 
\hline
\hline
\end{tabular}
\tablefoot{
\tft{a}{Velocity range over which the line was integrated.}
\tft{b}{Offset from the peak of continuum emission position.}
}
\end{table*}

Vibrational emission of HCN is excited by 14$\mu$m dust photons. In Arp~220 dust starts to be optically thick ($\tau\sim1$) at $\sim1$mm \citep{Downes2007,Sakamoto2008} toward the WN, with opacities increasing to $\tau\sim5.3$ and 1.7 at 434$\mu$m for the WN and EN, respectively \citep{Wilson2014}. Dust is optically thick toward both nuclei at 14$\mu$m.
Additionally, a large gas surface density is determined toward both nuclei \citep[$6\times10^{24}$ and $2\times10^{25} \rm cm^{-2}$, respectively,][]{Wilson2014}.
These conditions are ideal for the excitation of vibrational transitions, according to \citet{Aalto2015a}, and they are met in both nuclei.
In this case, the dust obscured IR emission is re-emitted through the submm and mm vibrational transitions of HCN, which gives us access to the inner regions of the nuclei
in which these transitions are excited.

However, in addition to the large amounts of material driving the high opacity, excitation like this needs high dust temperatures ($T_d>100$~K) for the emission at 14$\mu$m to be bright enough. This requirement is met in the western nucleus ($T_d\sim200$~K), but not in the eastern nucleus \citep[$T_d\sim80$~K,][]{Wilson2014}. We did not detect vibrational emission toward the EN, which means that there must exist a temperature gradient with temperatures $>100$~K at the center
of its opaque dust nucleus. A temperature gradient like this would be  responsible for the self-absorption that has been claimed to explain the line profiles
and kinematics of CO $J=6-5$ emission in the EN \citep{Rangwala2015}.
A similar gradient in the western nucleus would result in a temperature $>200$~K, which agrees with the $T_{\rm vib}>300$~K measured with HC$_3$N \citep{Mart'in2011}.

In Fig.~\ref{fig.vibmaps} we overplot the integrated intensity maps of the HCN $J=4-3~v_2=1f$ emission on the continuum emission. We used this transition because
it provides a higher signal-to-noise ratio and a slightly smaller synthesized beam than the corresponding $J=3-2$ transition maps.

Toward the EN we integrated the central $200$~\kms~ , corresponding to the velocity range of 440-640~\kms~(which we refer to as
the HCO$^+$ rest frequency, as displayed in Fig.~\ref{fig.lineprofiles}). We integrated the whole line down to the half-power since it is only moderately affected by blending on the blueshifted side.
Additionally, we imaged the vibrational emission in the blueshifted velocity range from $-100$ to 0~\kms~ with respect to the center of the line ($440-540$~\kms~ in Fig.~\ref{fig.lineprofiles}) and redshifted from 0 to 100~\kms~($540-640$~\kms~ in Fig.~\ref{fig.lineprofiles}). 
Gaussian fits were performed to these integrated emission maps. Results are summarized in Table~\ref{tab.vibimagefit}.
Although the continuum emission has a deconvolved size of $\sim140\times120$~pc, the HCN vibrational emission is confined to a barely resolved region of $\sim60\times50$~pc in the $4-3$ transition.
The fit to the $3-2$ vibrational transition with a slightly larger beam is consistent with a point source.
Vibrational emission is centered on the peak of the continuum. 
Similarly, the blue- and redshifted emissions show the same extent within the uncertainties and are also centered on the nuclear region.
At our resolution of $\sim0.5''$ ($\sim185$~pc) we cannot resolve any rotation around the nucleus on the HCN $v_2=1f$ line.
The large observed line width in the vibrational HCN emission suggests that it originates in the innermost regions of the EN. Based on the position and compactness
of the integrated emission and on the velocity of the emission (Sect.~\ref{sect.vibresults}), we can infer that we reached the regions at whose centers the mid-IR opaque dust originates.

On the other hand, the HCN $v_2=1f$ line is more severely affected by blending in the WN. We therefore only integrated the redshifted emission in the range of $20-170$~\kms.
Again, the extent of the deconvolved Gaussian fitted emission is $\sim60\times50$~pc, compared to the $\sim120\times60$~pc fitted to the continuum emission.
In this case, we observed a significant offset of $0.045''$ of the spectral feature toward the northwest (P.A.$\sim120^\circ$) with respect to the continuum, in good agreement with the
plane of rotation measured with HCN \citep[$P.A.\sim45^\circ$][]{Scoville2015}, but not with the gradient observed with CO in the inner region \citep{Sakamoto1999,Rangwala2015}.
Again, when peering through the opaque mid-IR dust, the HCN $v_2=1f$ line might be probing the kinematics of a hot dusty circumnuclear disk, as has been proposed by \citet{Aalto2015a}, with a radius of $\sim18$~pc.



\subsection{High $L_{vib}/L_{FIR}$ ratio}
\label{sect.lviblfir}

In Table~\ref{tab.vibfit} we include the line luminosity \citep[calculated follwoing Eq.1 in ][]{Solomon2005} of the two vibrationally excited lines measured in the two nuclei. We also present their ratios relative to the 
total luminosity of each nucleus as estimated by \citet{Wilson2014}.
The difference with the ratio reported by \citet{Aalto2015a} toward the WN arises because they assumed $log(L_{FIR})=12~L_\odot$.
We show the dependence of this ratio on the quantum number $J$, where the line luminosity ratio is $4-3/3-2\sim2.7$. This difference is similar to that of 2.5 derived in NGC~4418 \citep{Aalto2015a}.

The small sample of obscured galaxies compiled by \citet{Aalto2015a} allowed them to tentatively suggest that the ratio between the luminosity of the vibrationally excited line and the galaxy bolometric luminosity (or $L_{FIR}$ for these FIR-dominated
objects) might be inversely correlated with fast outflows. A relatively high vibrationally excited emission is therefore suggested as a probe of rapidly evolving nuclear growth.

The two nuclei of Arp~220 show the highest $L_{vib}/L_{FIR}$ ratio of all galaxies in which vibrationally excited emission has been detected so far.
Moreover, of the two nuclei, the EN has a relatively brighter luminosity in the vibrational transitions, where its ratio to the total luminosity is a factor of 2-3 higher (depending on the
way the vibrational emission is fit, as discussed in Sect.~\ref{sect.vibresults}) than what is measured toward the WN.
If we consider the more complex system of absorbers in the WN as a sign of an evolved nucleus compared to the single absorbing system in the EN, the lower $L_{vib}/L_{FIR}$ toward the WN compared to the EN would fit into this suggested scenario and relate the high luminosity ratio with the evolving nuclear growth.
However, this picture is only tentative and needs to be tested
in more detail.

Whether it is a quickly evolving supermassive black hole or a very compact starburst  giving rise to the bright emission in the vibrationally excited transitions \citep{Andrews2011,Aalto2015a}, it affects both nuclei. Toward the EN, where no evidence has been found of an embedded AGN, the vibrational emission would more likely originate in a hot starburst.
However, even a very compact starburst all filled with hot cores similar to Sgr~B2 within the central few hundred parsecs \citep{Cernicharo2006,Mart'in2011} would not be able to explain such efficient emission or pumping of the vibrational levels \citep{Aalto2015a} because we measure a luminosity ratio a few tens higher than what is measured in Galactic hot cores.

\section{Summary and conclusions}

We presented high-sensitivity and high-resolution ($\sim180$~pc) imaging of HCN and HCO$^+$ in the $3-2$ and $4-3$ transitions together
with the corresponding vibrationally excited transitions of HCN and the underlying continuum emission.

 HCN and HCO$^+$ are optically thick, and in addition, these lines are affected by various absorption systems and significant
line blending. These effects seem to be common to other compact obscured nuclei \citep{Aalto2015a}.
In Arp~220 we observed that absorptions alone could hide up to $\sim70\%$ of the total intrinsic HCN and HCO$^+$ emission lines, which may result in a poor tracing of the total content of dense gas in the most deeply buried regions, where an important fraction of the luminosity is generated. Therefore it could result in an artificial enhancement of the SFE in the most compact LIRGs and ULIRGs \citep{Gracia-Carpio2008}.
Moreover, these transitions may only be poorly suited for studying the kinematics of the molecular gas in the innermost regions
around the Arp~220 nuclei, and they might explain the discrepancy found in the moment maps of HCN and CS \citep{Scoville2015}, where the
absorption will have a different impact on each species.
Thus, these commonly used dense gas tracers severely limit studying the innermost
regions of compact obscured galaxy nuclei.

One of the key results from our observations is the clear detection of vibrationally excited HCN in the $4-3$ and $3-2$ transitions
not only toward the WN (previously reported by \citet{Aalto2015a}), but also toward the EN.
The detection of this emission strongly supports the scenario
of a hot dust component ($T_d>100$~K) within each of the two nuclei, which 
would require a temperature gradient toward the EN, where the measured dust temperature is $\sim 80$~K. Although the more
compact WN shows an observed dust temperature reaching $\sim 200$~K, which is enough for an efficient radiative pumping, it will most likely show a similar temperature gradient up to $>300$~K as revealed by the vibrational temperature measured with
HC$_3$N \citep{Mart'in2011}.

Vibrationally excited HCN traces regions much closer to the nucleus than those probed by the heavily absorbed ground-vibrational HCN and HCO$^+$ transitions.
In both nuclei we marginally resolved the vibrational emission extent to a region of
$60\times50$~pc, which is smaller than the measured continuum emitting regions of $\sim140\times120$~pc and $\sim120\times60$~pc in the EN and WN, respectively.
 
The excitation temperature derived between the vibrationally excited levels of $\sim 40$~K probably reflects the 
level population of the ground state of HCN \citep[similar to that measured for HC$_3$N,][]{Mart'in2011}, which is then efficiently pumped radiatively to the $v_2=1$ state through a strong $14\mu$m radiation field.
However, the peak flux density ratio between the two vibrational transitions might indicate that the lines are optically thick. In this case, opacity would again prevent us from studying these vibrational lines, and we would be limited to the external layers of the hot dust region that excites these transitions.
If these vibrational lines are optically thick, it would imply that they are being emitted from a much smaller region. From the optically thick limit we can establish lower limits
to the emitting region of 15~pc (EN) and 22~pc (WN). Even in this limit, the probed scales with HCN $v_2=1f$ would be much smaller than those for the ground-vibrational state transitions.
The kinematics of the vibrational emission can only be marginally hinted at with our current
resolution.

Based on the emission of HCO$^+~J=3-2,$ we measured the receding velocity of the individual nuclei of $V_{WN,LSR,radio}=5342\pm4$~\kms~ and $V_{EN,LSR,radio}=5454\pm8$\kms.
Vibrational emission is centered on these nuclear velocities within the uncertainties, again supporting the fact that these transitions probe the innermost regions around the nuclei.
Thus, these emission lines are unique probes of the nuclear enclosed masses.
Estimated virial masses are $M_{vir}^{WN}\sim6-9\times10^8~M_\odot$ and 
$M_{vir}^{EN}\sim2-3\times10^8~M_\odot$ within a radius of $25-30$~pc of each nucleus.
This results in gas surface densities of $3\pm0.3\times10^4~M_\odot~\rm pc^{-2}$ (WN) and $1.1\pm0.1\times10^4~M_\odot~\rm pc^{-2}$ (EN) for a conservative gas fraction $f_g=0.1$. Higher gas fractions would result in values close to the highest
stellar surface density of $\sim10^5~M_\odot~\rm pc^{-2}$ \citep{Hopkins2010}.


In both nuclei, we observed a high ratio between the vibrational line luminosity and the galaxy luminosity ($L_{\rm vib}/L_{FIR}$).
Although the ratio observed toward the EN is twice as high as that toward the WN, the ratio in both nuclei is much higher than any other ratio measured toward
other luminous galaxies \citep{Aalto2015a}. A relative vibrational emission this high is difficult to reconcile with a compact starburst origin with Galactic type hot cores.


The absorbing gas was found to be blueshifted at a velocity of $-50$~\kms~ relative to the EN velocity, while the WN shows absorptions at $6$ (which is consistent within the errors with the WN velocity), $-140$, and $-575$~\kms. 
The absorbing gas systems toward the EN were found to be blueshifted by $\sim -50$~\kms, and three systems in the WN shifted by -6, -140, and -575~\kms. These velocities refer to the individual velocities of each nucleus.
The blueshifted absorption systems may be associated with the outflowing clumps.
The blueshifted absorptions toward the WN were found to be marginally offset from the center by $\sim 20$~pc, which is consistent with the bipolar outflow scenario suggested for the SiO observations by \citet{Tunnard2015}.

We found that most ($\sim90\%$) of the continuum emission at 1.2 and 0.85~mm is associated with the two nuclei, which is consistent with the previous findings compiled in \citet{Matsushita2009}.

At the current resolution, we can infer some of the physical properties within the Arp~220 nuclei, but deeper insight will
be achieved through modeling of the vibrationally excited emission in compact obscured nuclei (Gonz\'alez-Alfonso in preparation) and
even higher spatial resolution observations of these infrared pumped transitions (Sakamoto in preparation).

\begin{acknowledgements}
This paper makes use of the following ALMA data: ADS/JAO.ALMA\#2012.1.00453.S. ALMA is a partnership of ESO (representing its member states), NSF (USA), and NINS (Japan),
together with NRC (Canada) and NSC and ASIAA (Taiwan), in cooperation with the Republic of Chile. The Joint ALMA Observatory is operated by ESO, AUI/NRAO, and
NAOJ.
S.A. acknowledges support from the Swedish National Science Council grant 621-2011-4143.
K.S. was supported by MOST grant 102-2119-M-001-011-MY3.
E.G.-A. is a Research Associate at the Harvard-Smithsonian
Center for Astrophysics, and thanks the Spanish Ministerio de Econom\'{\i}a y Competitividad for support under project FIS2012-39162-C06-01 and NASA grant ADAP NNX15AE56G.
F.C. acknowledges support from the Swedish National Science Council grant 637-2013-7261.
J.M.-P. acknowledges partial support by the MINECO under grants AYA2010-2169-C04-01, FIS2012-39162-C06-01, ESP2013-47809-C03-01 and ESP2015-65597-C4-1. 
\end{acknowledgements}

\bibliographystyle{aa}  
\bibliography{ALMAHCNHCOp.bib}  


\appendix
\section{Velocity-integrated emission, PV diagrams, and multi-Gauss fit to the spectra}
\begin{figure*}
\centering
\includegraphics[width=\linewidth]{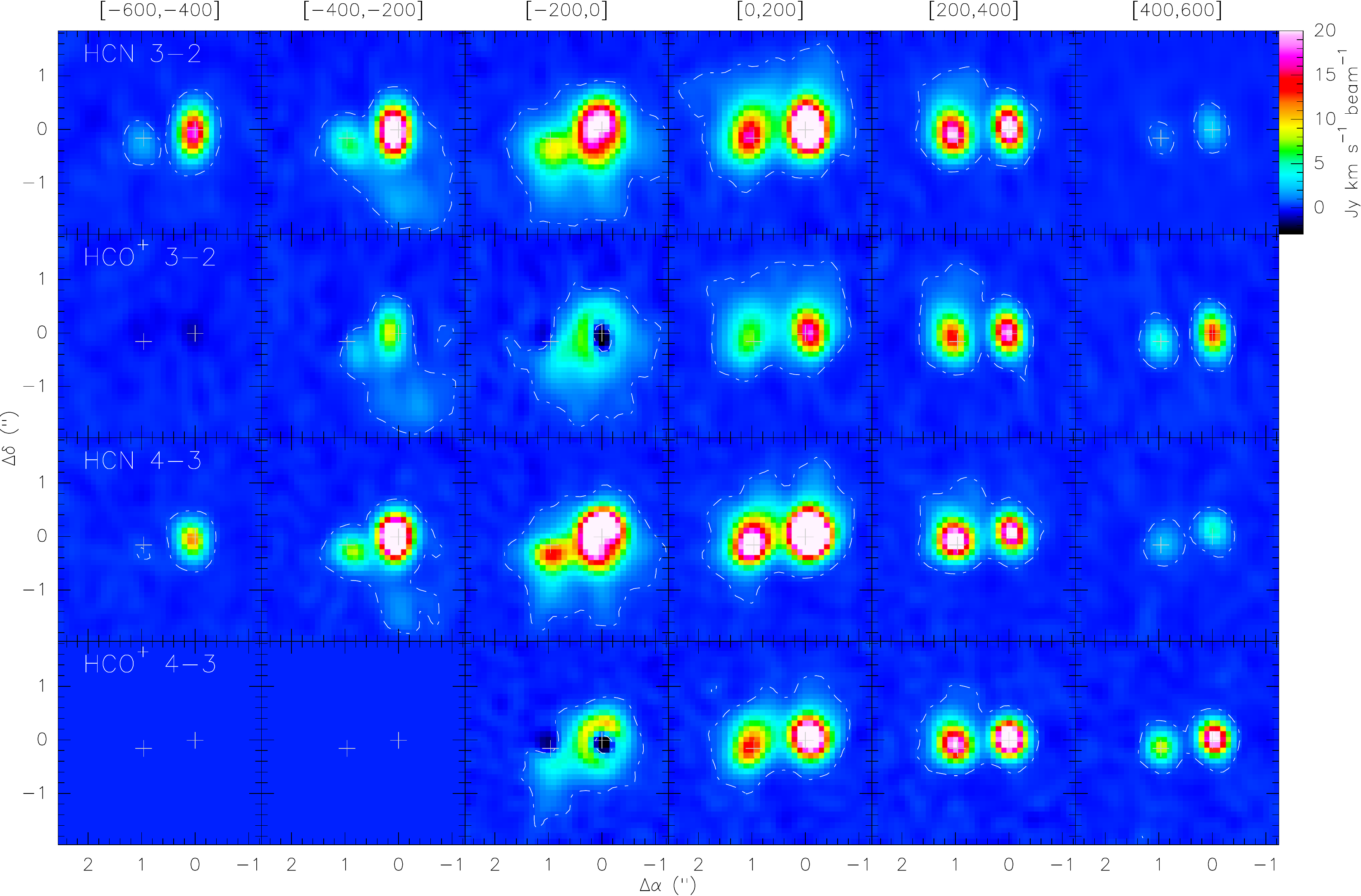}
\caption{Integrated intensity maps of HCN $3-2$, HCO$^+~3-2$, HCN $4-3$, and HCO$^+~4-3$ in 200~\kms bins from -600 to 600~\kms around the systemic velocity (see Sect.~\ref{Sect.HCNHCOp}). The single dot-dashed contour represents $3~\sigma$ ($1~\sigma=0.15$~mJy~\kms~beam$^{-1}$) and aims to guide the eye toward the most extended emission. Continuum emission peaks are indicated by gray crosses.
 \label{fig.integratedranges}}
\end{figure*}

\begin{figure*}
\centering
\includegraphics[width=\linewidth]{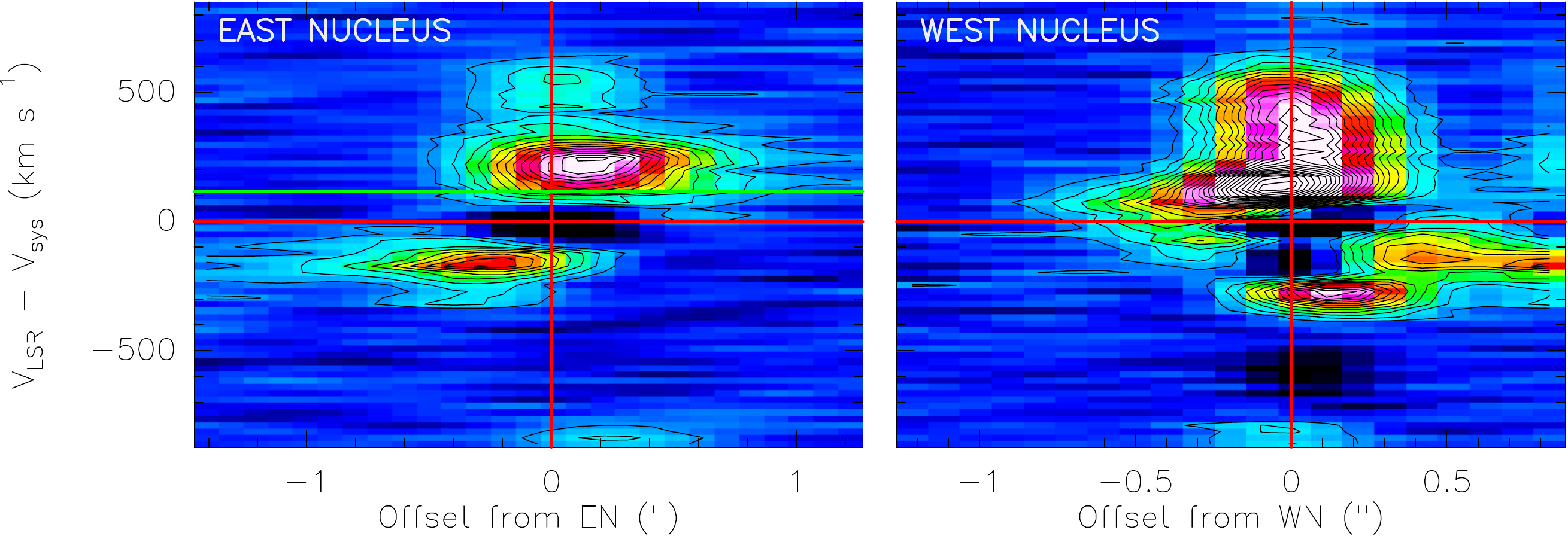}
\caption{Position-velocity diagrams across the two nuclei for the HCO$^+$ $3-2$ transition. The direction of each cut is shown by blue arrows in Fig.~\ref{fig.integrated}. The high-velocity emission extending
up to above 500~\kms~ corresponds to the vibrational emission of HCN (Fig.~\ref{fig.lineprofiles}).
Red lines show the V$_{\rm sys}$ of the entire system and the position of the nuclei. Toward the EN, the green line 
shows the derived velocity of the nucleus $V_{EN}-V_{sys}\sim117$~\kms.
\label{fig.pvdiagrams}}
\end{figure*}

\begin{figure*}
\centering
\includegraphics[width=\linewidth]{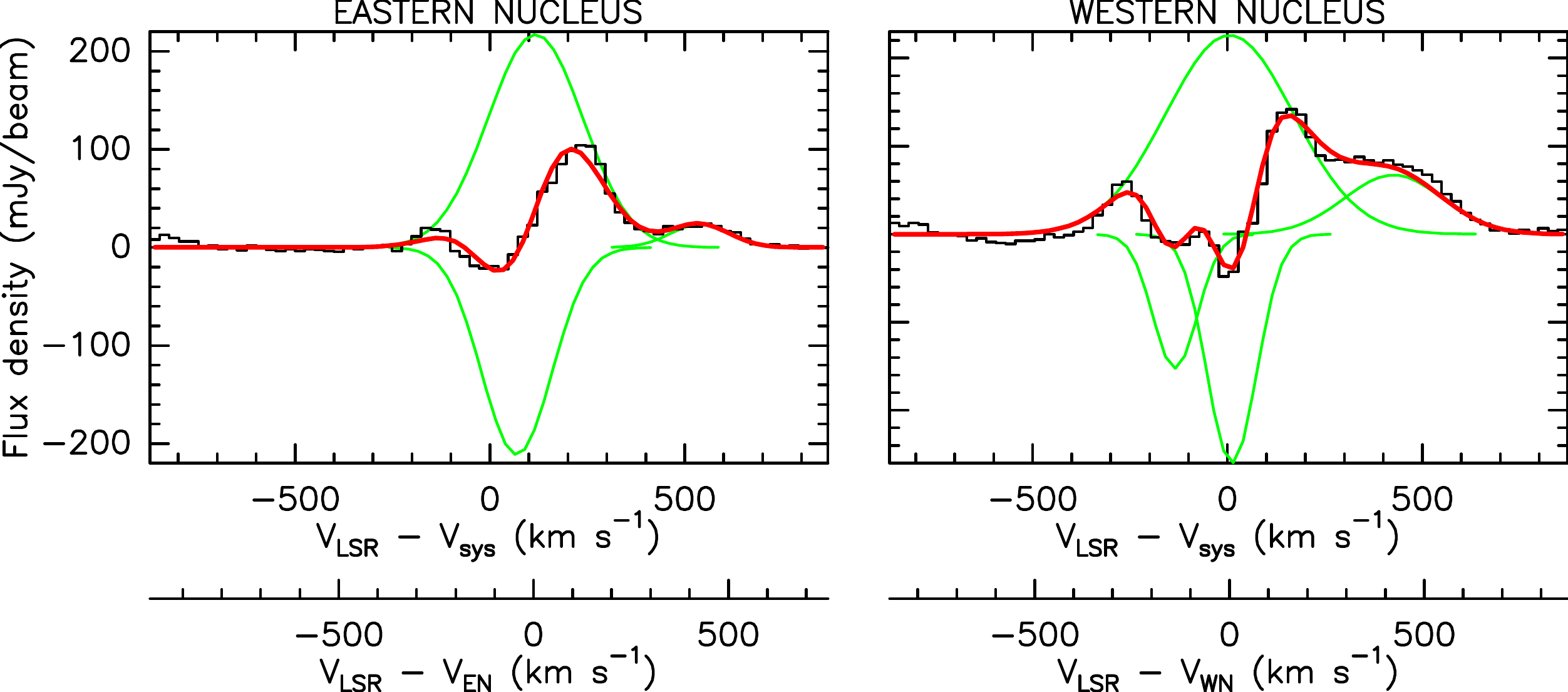}
\caption{Gaussian fit results to both the emission and absorption toward both nuclei as described in Sect~.\ref{sect.abssystems}.
In green we represent each emission and absorption component, where the emission from vibrational HCN emission is included, while in red we show the added profile.
The velocity scale is presented relative to the systemic velocity ($V_{sys}$) and to the derived individual velocity of each nucleus ($V_{EN}$ and $V_{WN}$, see Sect~\ref{sect.vsys}).
\label{fig.specabs}}
\end{figure*}

\end{document}